\documentstyle[preprint,aps,tighten]{revtex}

\begin{document}
\draft
\title{Application of a new screening model to thermonuclear reactions of the rp
process}
\author{Theodore E. Liolios \thanks{%
www.liolios.info}}
\address{Hellenic Naval Academy of Hydra\\
School of Deck Officers, Department of Science\\
Hydra Island 18040, Greece}
\maketitle

\begin{abstract}
A new screening model for astrophysical thermonuclear reactions was derived
recently which improved Salpeter's weak-screening one. In the present work
we prove that the new model can also give very reliable screening
enhancement factors (SEFs) when applied to the rp process. According to the
results of the new model, which agree well with Mitler's SEFs, the screened
rp reaction rates can be, at most, twice as fast as the unscreened ones.
\end{abstract}


In a recent work\cite{lioliosnew} a new model was derived for weakly
screened (WS) thermonuclear reactions, improving Salpeter's one\cite
{salpeter}. The screening enhancement factor (SEF) of the novel model was
shown to coincide with Salpeter's and Mitler's\cite{mitler} for
proton-proton solar reactions, thus confirming again that, as far as
screening uncertainties are concerned, the relevant pp neutrino fluxes are
confined within very robust limits.

Actually Salpeter's model suffers from the break-down of the WS limit (the
very limit that generates it) inside the tunneling region. On the other hand
Mitler's model has made the arbitrary assumption that at close distances
from all nuclei in the plasma the electron density is practically equal to
the average electron density in the plasma. The new model modified Mitler's
method assuming a very natural behavior for the charge density around the
nucleus and adopting the DH\ formalism only where it is really valid.

We briefly give the most essential information of the three models
(Salpeter's, Mitler's and the author's) for a binary proton-induced
thermonuclear reaction $\,p+_{Z}^{A}M_{N}\,$\thinspace \thinspace in
completely ionized astrophysical plasmas at temperature $T_{6}\,\,$and
density $\rho $ :

A) Salpeter's model

Salpeter's SEF is given by

\begin{equation}
f_{S}=\exp \left( \frac{Ze^{2}}{R_{D}kT}\right) =\exp \left( 0.188Z\chi \rho
^{1/2}T_{6}^{-3/2}\right)  \label{salp}
\end{equation}
where $R_{D}$ is the Debye-Huckel radius and $\chi \,$\thinspace the
parameter which $\,$incorporates all the information about plasma
composition and degeneracy:

\begin{equation}
\chi =\sqrt{\sum_{i\neq e}\frac{X_{i}Z_{i}^{2}}{A_{i}}+\theta \left(
a\right) \sum_{i\neq e}\frac{X_{i}Z_{i}}{A_{i}}}
\end{equation}
Electron degeneracy is taken into account via the degeneracy factor$\,$%
\thinspace $\theta \left( a\right) \,\,$which is a function of the
respective degeneracy parameter $a\,$(for details see \cite{lioliosnew}),
while $X_{i}$ is the abundance of nuclei $_{Z_{i}}^{A_{i}}M\,\,$in the
plasma.

B) Mitler's model

According to Mitler's model at short distances from all nuclei the electron
density is constant and equal to the average electron density in the plasma.
The arbitrariness of this assumption is obvious since it can only apply
safely to completely degenerate electron environments. Even in strongly
degenerate plasmas a proton-rich nucleus is expected to polarize the
electron cloud in its immediate vicinity thus rendering the assumption of a
constant electron density questionable. The degree to which this
polarization occurs is actually the degree of the error committed by
Mitler's model. Therefore, when thermonuclear reactions of the rp-process
are considered and the degeneracy is usually incomplete Mitler's assumption
should be questioned.

The distance from the target nucleus at which the constant electron density $%
N_{e}\,\,$is replaced by the Debye-Huckel density, according to Mitler is:

\begin{equation}
x=\left( \zeta +1\right) ^{1/3}-1
\end{equation}
where the dimensionless parameter $\zeta $ is:

\begin{equation}
\zeta \left( Z,\rho ,T\right) =\frac{3Z}{4\pi N_{e}R_{D}^{3}}  \label{zdms}
\end{equation}
In the framework of the same model the screening energy shift $U_{e}^{M}$ is

\begin{equation}
U_{e}^{M}=\frac{Z_{1}Z_{2}e^{2}}{R_{D}kT}\left( \frac{1+x/2}{1+x+x^{2}/3}%
\right)  \label{uem}
\end{equation}
which yields Mitler's SEF :

\begin{equation}
f_{M}=\exp \left[ \frac{Ze^{2}}{R_{D}kT}\left( \frac{1+x/2}{1+x+x^{2}/3}%
\right) \right]  \label{sefmitsl}
\end{equation}
We can vividly depict the connection between Mitler's SEF and its generator
(i.e. Salpeter's SEF) by writing

\begin{equation}
f_{M}=\left( f_{S}\right) ^{g\left( x\right) }  \label{sefm1}
\end{equation}
where

\begin{equation}
g\left( x\right) =\left( \frac{1+x/2}{1+x+x^{2}/3}\right)  \label{gsl}
\end{equation}
Mitler also elaborated his model considering the effects of the cloud of the
second member $\left( Z_{2}e\right) \,$of the binary thermonuclear reaction
and deriving a formula for the relevant SEF which, after some algebra, can
be written again as a function of Salpeter's SEF:

\begin{equation}
f_{M}^{*}=\left( f_{S}\right) ^{g^{*}\left( \zeta _{1},\zeta _{2}\right) }
\label{sefm2}
\end{equation}
where

\begin{equation}
g^{*}\left( \zeta _{1},\zeta _{2}\right) =\frac{9}{10}\left( \frac{1}{\zeta
_{1}\zeta _{2}}\right) \left[ \left( \zeta _{1}+\zeta _{2}+1\right)
^{5/3}-\left( \zeta _{1}+1\right) ^{5/3}-\left( \zeta _{2}+1\right)
^{5/3}+1\right]
\end{equation}
and $\zeta _{1},\zeta _{2}$ are the dimensionless parameters given by Eq.$%
\,\left( \ref{zdms}\right) $ for each of the two reacting nuclei. Since we
will study proton-induced reactions let us now assume that the parameter $%
\zeta _{2}$ corresponds to a proton then $\zeta _{1}=Z\zeta _{2}$ where we
have dropped the indices from the atomic number so that $Z_{1}=Z$ and $%
Z_{2}=1.\,$\thinspace In such a case

\begin{equation}
g^{*}\left( Z,\rho ,T\right) =\frac{9}{10}\left( \frac{1}{Z\zeta ^{2}}%
\right) \left[ \left( Z\zeta +\zeta +1\right) ^{5/3}-\left( Z\zeta +1\right)
^{5/3}-\left( \zeta +1\right) ^{5/3}+1\right]  \label{gcomplex}
\end{equation}

C) The novel model.

In Ref. \cite{lioliosnew} a new model was derived which avoids any arbitrary
assumption about the electron density around the target nucleus. According
to that model the SEF for a proton-induced thermonuclear reaction $%
p+_{Z}^{A}M_{N}$ \thinspace should be given as a function of Salpeter's SEF
modified as follows

\begin{equation}
f=f_{s}^{G\left( x_{0},x_{0}^{^{\prime }}\right) }  \label{mysef}
\end{equation}
where the parameters $x_{0},x_{0}^{^{\prime }}$ are obtained by solving the
following equations:

\begin{equation}
\frac{e^{x_{0}^{^{\prime }}}}{x_{0}^{^{\prime }3}}\left[
2-e^{-x_{0}^{^{\prime }}}\left( x_{0}^{^{\prime }2}+2x_{0}^{^{\prime
}}+2\right) \right] =\frac{e^{x_{0}}}{x_{0}^{2}}\left[ 1-e^{-x_{0}}\left(
x_{0}+1\right) \right]  \label{final1}
\end{equation}

\begin{equation}
x_{0}e^{x_{0}}=1.88Z_{\max }Z\chi \rho ^{1/2}T_{6}^{-3/2}  \label{x0}
\end{equation}
and the exponent is given by the relation

\begin{equation}
G\left( x_{0},x_{0}^{^{\prime }}\right) =\frac{1}{x_{0}}-\frac{e^{-x_{0}}}{%
x_{0}}-2\frac{x_{0}}{x_{0}^{^{\prime }3}}e^{x_{0}^{^{\prime }}-x_{0}}+\frac{%
x_{0}}{x_{0}^{^{\prime }3}}\left( 2+x_{0}^{^{\prime }}\right) e^{-x_{0}}+%
\frac{x_{0}}{x_{0}^{^{\prime }2}}e^{x_{0}^{^{\prime }}-x_{0}}  \label{gx0}
\end{equation}
Actually, Eq. $\left( \ref{gx0}\right) \,\,$appears in the derivation of a
novel screened Coulomb potential $\Phi \left( r\right) \,$\thinspace given by%
\cite{lioliosnew}

\begin{equation}
\Phi \left( r\right) =\frac{Z_{0}e}{r}-\frac{Z_{0}e}{R_{D}}G\left(
x_{0},x_{0}^{^{\prime }}\right) +O\left( r^{2}\right)
\end{equation}
That potential, by means of a well established mechanism\cite{lioliosprc2000}%
, yields a shift in the relative energy (the screening energy) which reads

\begin{equation}
U_{e}=\frac{Z_{0}Z_{1}e^{2}}{R_{D}}G\left( x_{0},x_{0}^{^{\prime }}\right)
\label{ueg}
\end{equation}
Using Eq. $\left( \ref{ueg}\right) \,$in the framework of Salpeter's model
described above, we arrive at Eq. $\left( \ref{mysef}\right) \,\,$

According to Ref. \cite{lioliosnew}, for a pure hydrogen plasma $Z_{\max }=1$%
,\thinspace while for a zero metallicity plasma $Z_{\max }=2.$ The fact that
the novel model disregards all screening effects of nuclei other than
protons and alpha particles can be easily justified since for stellar
environments where the rp-process takes place the abundances of nuclei other
than protons and alpha-particles are orders of magnitude smaller. Therefore,
as regards composition, the parameter $\chi \,\,$appearing in Eq. $\left( 
\ref{x0}\right) \,\,$is practically an exclusive function of $%
X_{1},X_{2},A_{1},A_{2},Z_{1},Z_{2}\,,\,$which justifies the zero
metallicity scenario.

We need to underline that the above three models A,B,C are valid provided
that all nuclei in the plasma are in a non-degenerate liquid state. In a
stellar environment where the rp process occurs that condition is fully
satisfied even at ultra-degenerate conditions such as $T=10^{6}\,K$ and $%
\rho =10^{6}\,g/cm^{3}.$

To prove the reliability of the new model we will first show that screening
effects in the rp-process follow a very simple pattern. The charge carried
along by the proton as it collides with a proton-rich target can be
disregarded and so can cloud fluctuations and dynamic effects. This can be
proved in the following plausible way:

In proton-induced thermonuclear reactions of the form\thinspace $%
\,p+_{Z}^{A}M_{N}$ \thinspace we can define two possible limits according to
those defined in the laboratory\cite{liolioslab2001,liolioslab2002}:

I) The Sudden Limit (SL), where the charge density around the target nucleus
remains unaffected by the presence of the impinging proton, throughout the
tunneling process. (In pycnonuclear reactions that limit was called ''static
lattice approximation''\cite{svh})

II) The Adiabatic Limit (AL), where the charge density around the target
nucleus is assumed to respond so fast that it actually corresponds to the
charge $\left( Z+1\right) e$ of a new combined nucleus consisting of the
initial target nucleus $\left( Ze\right) $ plus the impinging proton $\left(
+e\right) $. (In pycnonuclear reactions that limit was called ''fully
relaxed approximation''\cite{svh})

These two assumptions bracket the behavior of the plasma screening effect so
that all other fine phenomena such as charge cloud deformations, dynamic
screening etc. are included in these two limits. Note that Mitler has
pointed out that the application of the target-projectile model on SEFs is
wrong (see for example Ref. \cite{cameron})\thinspace because the ensuing
formulas are not commutative with respect to the target-projectile pair as
demanded by the thermonuclear reaction rate. However, this is not the case
when protons react with heavily charged nuclei because the classical turning
point of such reactions is so deep inside the screening cloud of the target
nucleus that it makes no difference which member of the pair the cloud is
attributed to. Thus, this common cloud can be either considered completely
frozen during tunnelling (SL) or rapidly responding to the presence of the
proton (AL).

In Figure 1 we plot the variation of the exponent $g$ (given by Eqs. $\left( 
\ref{gsl}\right) ,\left( \ref{gcomplex}\right) $) with respect to the
parameter $\zeta $ (given by Eq.$\left( \ref{zdms}\right) )\,\,$for the
three different limits discussed in the text$.$ The parameter $\zeta $
assumes all its possible values, that is from very small values occurred in
weakly degenerate, weakly screened environments to large values occurred in
completely degenerate, strongly screened ones. The arrows indicate that
screening and degeneracy are both increasing functions of $\zeta $. The
solid curve corresponds to Mitler's Sudden Limit (i.e. Eq. $\left( \ref{gsl}%
\right) )$, the dotted curve corresponds to Mitler's Adiabatic Limit (i.e.
setting $Z\rightarrow Z_{1}+1\,$in Eq. $\left( \ref{gsl}\right) )$ and the
dashed curve corresponds to Mitler's complex SEF which takes into account
fine screening effects (i.e. Eq.$\left( \ref{gcomplex}\right) )$. By
observing Fig. 1 we can easily realize that for reactions of the rp-process
where $Z>7$ all fine screening effects of Mitler's model can be disregarded.
Moreover the difference between the SL and the AL in such reactions is so
narrow that Mitler's model can be accurately represented in the rp-process
by the use of its simple SL SEF, i.e. Eq. $\left( \ref{sefm1}\right) $.

Note that the method of the SL and AL should be applied with caution to the
CNO solar cycle where even a small perturbation in the value of the SEF can
cause notable uncertainties to the neutrino fluxes\cite{bahcallbook,ricci}.

In order to illustrate the degree of validity of the three models when
applied to the solar CNO cycle we produced Figure 2 (in accordance with
Figure 3 of Ref. \cite{dzitko}). According to that figure the discrepancy
between the SL and the AL\ observed for both Mitler's and the author's
models is large enough to cause notable uncertainties in the production of
the solar neutrino fluxes. According to Fig.2 the SEF for the solar reaction 
$^{14}N\left( p,\gamma \right) ^{15}O$ \thinspace is confined within a
maximum $\left( f_{\max }\right) $ and a minimum $\left( f_{\min }\right) $
\thinspace value derived respectively by the author's AL model and Mitler's
complex model. The two values bear a difference of 7\%\thinspace , which
reflects linearly on the neutrino fluxes generated by the reactions $%
^{13}N\left( e^{+}\nu _{e}\right) ^{13}C\,\,$and $^{15}O\left( e^{+},\nu
_{e}\right) ^{15}N.\,\,$However, as has been pointed out in Ref. \cite
{lioliosnew} it is very reasonable to assume that the respective SEF cannot
practically lie outside the robust AL and SL limits of the novel model,
which deviate from each other by 3.5\%.

Although the new model allows for the marginal uncertainty of 3.5\% in the
CNO cycle, when applied to more advanced stages such as the rp process that
uncertainty becomes negligible.

In figures 3a,3b,3c we plot the plasma SEFs with respect to density for a
very important reaction\cite{walwoo} of the rp-process\thinspace $%
^{18}F\left( p,\gamma \right) ^{19}Ne$ \thinspace according to the three
models described in the text at temperatures $T_{6}=100,\,500,\,1000$ . The
density $\rho _{2}\,\left( =\rho /100\,g/cm^{3}\right) \,\,$ranges from
typical solar values to extremely large ones typically found on neutron star
surfaces. For simplicity we have assumed a zero-metallicity stellar plasma
and a hydrogen-helium composition typical\cite{walwoo} of the rp-process : $%
X=0.7,Y=0.3.\,$We have also added Salpeter's screening formula\cite{salpeter}
for completely degenerate plasmas which is actually the limit of Mitler's
model for similar conditions. We observe that for plasmas which are not
completely degenerate the author's model gives roughly the same results as
Mitler's one. However, for ultradegenerate (UD) environments (irrelevant to
the rp process) Mitler's formula is more reliable than the author's as it
approximates better the relevant Salpeter's SEF formula

\begin{equation}
f_{s}^{UD}=0.205\left( \frac{\rho }{\mu _{e}}\right) ^{1/3}\left[ \left(
Z+1\right) ^{5/3}-Z^{5/3}-1\right]  \label{ssef}
\end{equation}
whose validity is very plausible in pycnonuclear regimes\cite{svh}.

In Figures 4a,4b we derive the SEFs for another important reaction of the
rp-process $^{50}Fe\left( p,\gamma \right) ^{51}Co$ \thinspace at relevant
temperatures. For such proton-rich targets as $^{50}Fe\,\,$the author's
model is much more reliable than Mitler's whose assumption that the electron
density around the target nucleus $^{50}Fe$ cannot practically be accurate.
The coincidence of the three models in question is due to the fact that at
large temperatures the plasma is only weakly screened and therefore even
Salpeter's model gives reasonable results.

We should point out that throughout this study we have focused on binary
thermonuclear reactions disregarding all nuclear correlation effects which
introduce\cite{ichimaru} an additional (multiplicative) screening
enhancement factor to the thermonuclear reaction rate. In fact, in the
density-temperature domain of the rp process the stellar plasma is only
weakly coupled (see for example Fig.1 of Ref. \cite{svh}), which means that
the Coulomb energy between ions is much smaller than the average thermal
energy of the ionic fluid. For such weakly coupled plasmas all internuclear
many-particle correlations can be disregarded\cite{svh,ichimaru} since the
plasma coupling constant is always $\Gamma \ll 1$. 

We can also use the ratio\cite{ichimaru} $\Lambda $ \thinspace between the
thermal De Broglie wavelength and the ionic spacing to support our
simplification. This ratio for all reactions, temperatures and densities
encountered in the rp process is much smaller than unity $\left( \Lambda \ll
1\right) ,\,$which justifies the assumption that all microscopic nuclear
correlations can be disregarded (for a detailed discussion see Ref.\cite
{ichimaru}).

Conclusions

Naturally, Salpeter's model cannot be applied to thermonuclear reaction of
the rp process, since as has been proved in Ref.\cite{lioliosnew} its
validity can only be justified in the study of pp reactions of weakly
screened, weakly degenerate environments.

Mitler's model on the other hand assumes a constant electron density around
the target nucleus which cannot be taken for granted for proton-rich nuclei
and partially degenerate stellar environments. In fact Mitler's model
doesn't tend to Salpeter's formula for completely degenerate environments
where Salpeter's Eq. $\left( \ref{ssef}\right) $is valid. It actually tends
to it when the parameter $\zeta \,\,$is very large and that is an
imperfection that should be noted.

The new model (C)\ fortified with the SL and AL assumptions seems to be the
most reliable one for the description of the screening enhancement effect in
the rp process. The most important aspect of the new model is that it can
derive a SEF as a function of Salpeter's SEF which in turn can take into
account various plasma processes.

We have applied our model to various other rp astrophysical reactions and we
have observed that the screening effect can accelerate the thermonuclear
reaction rates of the rp process by (at most) a factor of two.

FIGURE CAPTIONS

Figure 1. The variation of the exponent $g$ (given by Eqs. $\left( \ref{gsl}%
\right) ,\left( \ref{gcomplex}\right) $) with respect to the parameter $%
\zeta $ (given by Eq.$\left( \ref{zdms}\right) )\,\,$for the three different
limits discussed in the text$.$ The parameter $\zeta $ assumes all its
possible values, that is from very small values occurred in weakly
degenerate, weakly screened environments to large values occurred in
completely degenerate, strongly screened ones. The arrows indicate that
screening and degeneracy are both increasing functions of $\zeta $. The
solid curve corresponds to Mitler's Sudden Limit (i.e. Eq. $\left( \ref{gsl}%
\right) )$, the dotted curve corresponds to Mitler's Adiabatic Limit (i.e.
setting $Z\rightarrow Z_{1}+1\,$in Eq. $\left( \ref{gsl}\right) )$ and the
dashed curve corresponds to Mitler's complex SEF which takes into account
fine screening effects (i.e. Eq.$\left( \ref{gcomplex}\right) )$.

Figure 2

The plasma SEF for the reaction $^{14}N\left( p,\gamma \right) ^{15}O$
\thinspace according to the three models described in the text at $T_{6}=15$
. For simplicity we have assumed a zero-metallicity stellar plasma and a
hydrogen-helium composition of : $X=0.7,Y=0.3.\,$The upper (lower) solid
curve stands for the author's AL (SL) model, the upper (lower) dashed curve
stands for Mitler's AL (SL) model, the dash-dotted curve represents Mitler's
formula for two ionic screening clouds while the dotted curve stands for
Salpeter's weak-screening model .

Figure 3a.

The plasma SEF with respect to density for the reaction $^{18}F\left(
p,\gamma \right) ^{19}Ne$ \thinspace according to the three models described
in the text at $T_{6}=100$ . For simplicity we have assumed a
zero-metallicity stellar plasma and a hydrogen-helium composition typical of
the rp-process : $X=0.7,Y=0.3.\,$The solid curve stands for the author's
model, the dashed curve stands for Mitler's SL model while the dotted curve
stands for Salpeter's weak-screening model. We have also added the
dash-dotted curve which corresponds to Salpeter's screening formula for
completely degenerate plasmas. The vertical bar indicates the barrier beyond
which electron degeneracy is complete.

Figure 3b.

The plasma SEF with respect to density for the reaction $^{18}F\left(
p,\gamma \right) ^{19}Ne$ \thinspace according to the three models described
in the text at $T_{6}=500$ (see figure 3a for details).

Figure 3c.

The plasma SEF with respect to density for the reaction $^{18}F\left(
p,\gamma \right) ^{19}Ne$ \thinspace according to the three models described
in the text at $T_{6}=1000$\thinspace \thinspace (see figure 3a for details).

Figure 4a

The SEFs for the reaction $^{50}Fe\left( p,\gamma \right) ^{51}Co$
\thinspace according to the three models described in the text at $%
T_{6}=300\,\,$(see figure 3a for details).

Figure 4b

The SEFs for the reaction $^{50}Fe\left( p,\gamma \right) ^{51}Co$
\thinspace according to the three models described in the text at $%
T_{6}=1000\,\,$(see figure 3a for details).

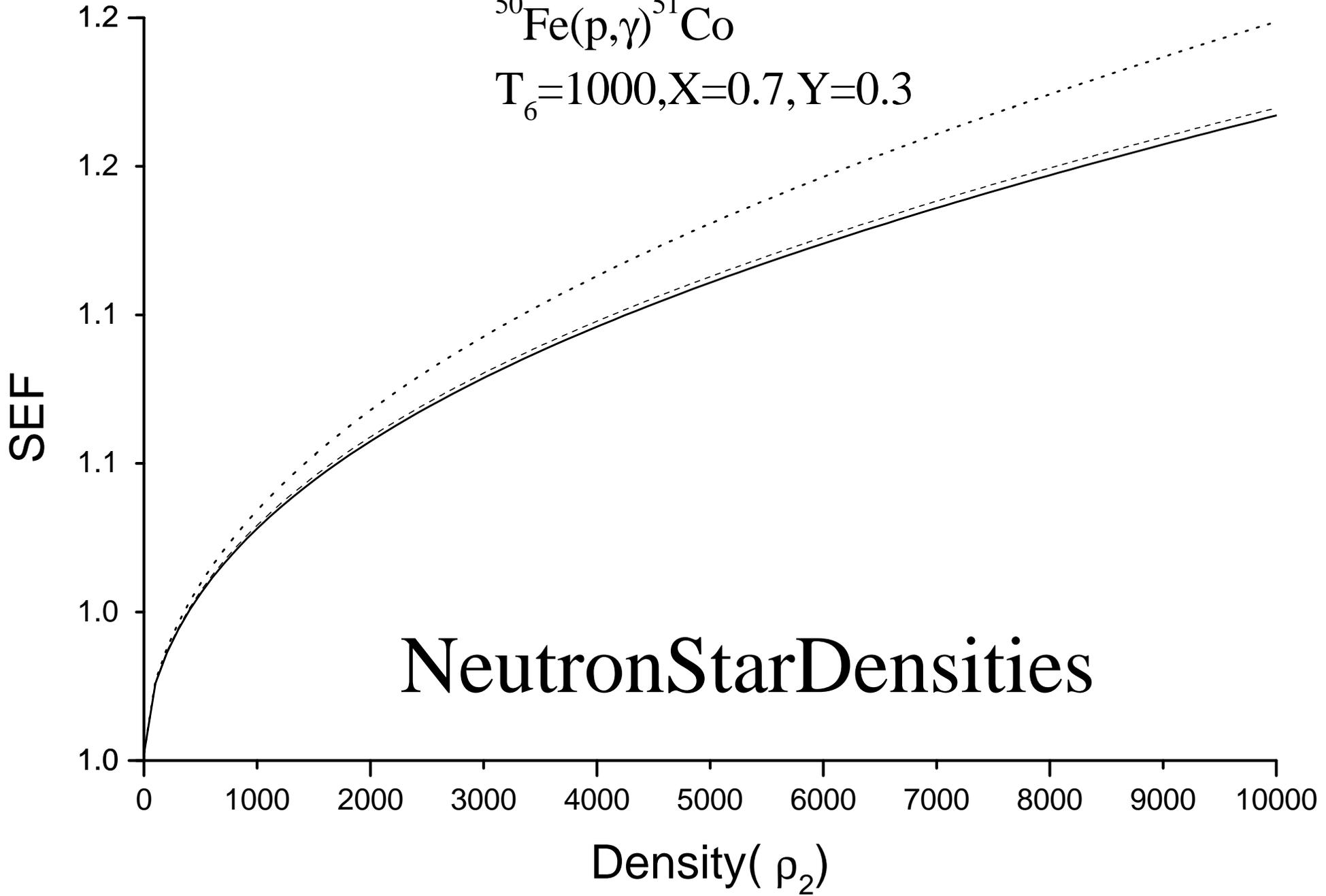
\end{document}